\documentclass[aps,prl,superscriptaddress,twocolumn,showpacs,floatfix]{revtex4}
\usepackage{graphicx}

\begin{document}


\title{Optical Production of Ultracold Polar Molecules}


\author{Jeremy M. Sage}
\author{Sunil Sainis}
\affiliation{Department of Physics, Yale University, New Haven, CT
06520, USA}
\author{Thomas Bergeman}
\affiliation{Department of Physics and Astronomy, SUNY, Stony
Brook, NY 11794-3800, USA}
\author{David DeMille}
\affiliation{Department of Physics, Yale University, New Haven, CT
06520, USA}


\date{\today}

\begin{abstract}
We demonstrate the production of ultracold polar RbCs molecules in
their vibronic ground state, via photoassociation of laser-cooled
atoms followed by a laser-stimulated state transfer process. The
resulting sample of $X ^1\Sigma^+ (v=0)$ molecules has a
translational temperature of $\sim\!100~\mu$K and a narrow
distribution of rotational states. With the method described here
it should be possible to produce samples even colder in all
degrees of freedom, as well as other bi-alkali species.

\end{abstract}


\maketitle

Samples of ultracold, polar molecules (UPMs) can provide access to
new regimes in many phenomena.  Ultracold temperatures allow
trapping, and polarity can be used to engineer large, anisotropic,
and tunable interactions between molecules.  These features make
UPMs attractive as qubits for quantum computation
\cite{DeMille02}, as building blocks for novel types of many-body
systems \cite{Baranov02}, and for the study of chemistry in the
ultracold regime \cite{Bodo02}.  Furthermore, UPMs can be used as
uniquely sensitive probes of phenomena beyond the Standard Model
of particle physics \cite{Kozlov95}.

Methods such as buffer-gas cooling \cite{Weinstein98},
Stark-slowing \cite{Bethlem00}, billiard-like collisions
\cite{Ellioff03}, and velocity filtering \cite{Rangwala02} have
produced samples of polar molecules at temperatures of
$\sim\!\!10$--$100$ mK. Formation of heteronuclear molecules from
pre-cooled atoms via photoassociation (PA)
\cite{Kerman04,Mancini04,Haimberger04,Wang04} or Feshbach
resonance techniques \cite{Inouye04} promises access to much lower
temperatures; however, these processes leave molecules in highly
excited vibrational levels, which have vanishingly small polarity
\cite{Kotochigova03} and are unstable to collisions
\cite{Yurovsky00,Mukaiyama04}.  The possibility of transferring
such molecules to their vibronic ground state, via optical
processes such as stimulated Raman transitions, has been discussed
extensively (see e.g.
\cite{DeMille02,Damski03,Bergeman04,Stwalley04,Kotochigova04});
however, insufficient data on the structure of experimentally
accessible molecules has made it difficult to identify specific
pathways for efficient transfer.

Here we report the production of UPMs via PA of laser-cooled Rb
and Cs atoms, followed by a two-step stimulated emission pumping
(SEP) process. This yields RbCs molecules in their absolute
vibronic ground state $X ^1\Sigma^+ (v=0)$.  These polar molecules
(calculated electric dipole moment $\mu \approx 1.3~\textrm{D}$
\cite{IgelMann86}) have a translational temperature of \(\sim \!
100~\mu\)K. The distribution of rotational states is also quite
narrow, so the resulting sample of $X ^1\Sigma^+$ state molecules
is cold in all degrees of freedom.

\begin{figure}
\centering
\includegraphics[width=3.4in]{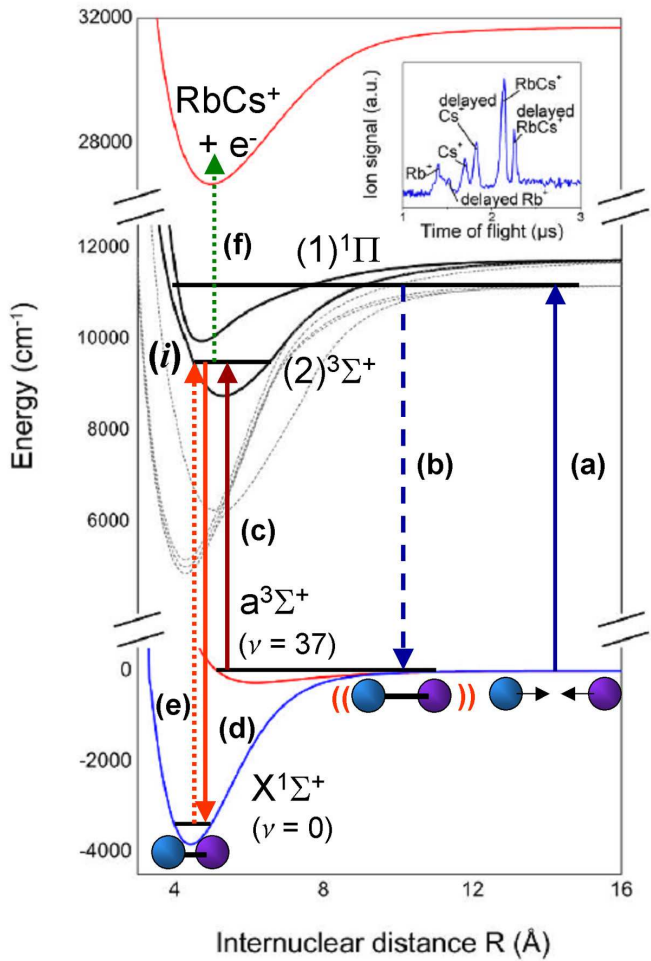}
\caption{(color online)  Formation and detection processes for
ultracold ground state RbCs. (a) Colliding atom pairs are excited
into weakly bound \(\mathrm{RbCs}^{*}\) molecules, which (b) decay
prominently into the \(a^{3}\Sigma^{+} (v=37)\) state.  (c)
Metastable \(a(v=37)\) molecules are excited to level \textit{i},
then (d) stimulated down into the \(X^{1}\Sigma^{+} (v=0)\) state.
The molecules are detected directly by (e) driving them back to
the original excited level and (f) ionizing them, or indirectly by
detecting the depletion of the $i$ level population (where only
(f) is needed). Inset:  a typical time-of-flight mass spectrum
showing the direct detection of \(X(v = 0)\) molecules.  The ion
signal, averaged over 200 shots, is plotted vs. delay time after
the ionizing pulse.
The delayed \(\mathrm{RbCs}^{+}\) peak signifies ground state
molecule production. } \label{fig01}
\end{figure}

     Figure \ref{fig01}
shows the methods by which we produce and detect UPMs.  A pair of
colliding, ultracold Rb and Cs atoms is photoassociated, i.e., the
pair absorbs a photon and is driven to an electronically excited
molecular level \cite{Thorsheim87}. This level decays rapidly,
with branching fraction of \( \sim\!7\)\% into the long-lived
\(a^{3}\Sigma^{+} (v=37)\) level \cite{Kerman04}. After a period
of PA, a resonant laser pulse (``pump'' pulse) transfers these
metastable, vibrationally excited molecules to an intermediate,
electronically excited state (\textit{i}). The population of state
\textit{i} is monitored by applying an intense laser pulse
(``ionization'' pulse), with a frequency chosen to selectively
form a \(\mathrm{RbCs}^{+}\) molecular ion. We detect these ions
using time-of-flight mass spectroscopy.

To produce \(X(v=0)\) molecules, a second tunable laser pulse
(``dump'' pulse), arriving just after the pump pulse, resonantly
drives molecules in state \textit{i} to the \(X(v=0)\) (or, for
diagnostic purposes, \(v=1\)) state.  Transfer to the \(X(v=0,1)\)
states is indicated by a depletion of the \textit{i} state
population.  This method \cite{Cooper81} generally works well, but
is complicated by the presence of additional resonant features
associated with transitions from state \textit{i} upward into
other, spectrally uncharacterized excited states.

We thus employ a second method to directly detect the RbCs
\(X(v=0,1)\) molecules.  After the dump pulse, the population of
state \textit{i} is allowed to decay for several times its
spontaneous emission lifetime, leaving essentially no population
in this level.  Next a third pulse (``re-excitation'' pulse),
identical in frequency to the dump, drives the stable
\(X^{1}\Sigma^{+}(v=0,1)\) molecules back into state \textit{i},
where they are detected via ionization as before.  The ion signal
is monitored as the frequency of the identical dump and
re-excitation pulses is scanned; a peak in this signal indicates
that molecules have been resonantly transferred to the
\(X^{1}\Sigma^{+}(v=0,1)\) states by the dump pulse, then back to
the \textit{i} state by the re-excitation pulse.  The definitive
signatures of UPM production are resonances at the exact
frequencies predicted by earlier spectroscopy of the
\(a^{3}\Sigma^{+}\), \textit{i} \cite{Kerman04,Bergeman04}, and
\(X^{1}\Sigma^{+}\) levels \cite{Fellows99}.

The state \textit{i} is in the manifold of levels associated with
the overlapping electronic states \(c^{3}\Sigma^{+}\),
\(b^{3}\Pi\), and \(B^{1}\Pi\). Specifically, the states chosen
for \textit{i} have predominantly \(c^{3}\Sigma^{+}_1\) character,
with small admixtures of \(b^{3}\Pi_1\) (due to non-adiabatic
couplings of the potentials) and \(B^{1}\Pi_1\) (due to spin-orbit
coupling). The mixed $c$/$B$ state composition of \textit{i} is
crucial to the technique (the small $b$ state admixture is
incidental). The \(c^{3}\Sigma^{+}\) component of the \textit{i}
state has reasonable Franck-Condon (FC) overlap with the initial
\(a^{3}\Sigma^{+} (v=37)\) state, due to the near-coincidence of
the inner turning point of the $a^{3}\Sigma^{+}$ potential with
the minimum of the \(c^{3}\Sigma^{+}\) potential
\cite{Kerman04,Bergeman04}.  The \(B^{1}\Pi\) component of state
$i$ circumvents the usual selection rule forbidding transitions
from the initial triplet state \(a\) to the final singlet state
\(X\). Moreover, the minima of the \(B^{1}\Pi\) and
\(X^{1}\Sigma^{+}\) potentials nearly coincide, leading to large
FC factors for transfer to the $X(v=0)$ level.

The apparatus is similar to that described in our earlier work
\cite{Kerman04}.  Briefly, \({}^{85}\mathrm{Rb}\) and
\({}^{133}\mathrm{Cs}\) atoms were cooled and collected in a
dual-species, forced dark SPOT magneto-optical trap (MOT)
\cite{Kerman04b,Ketterle93,Anderson94}.  The atomic density $n$
and atom number $N$ were \(n_{\mathrm{Rb}} = 1 \times
10^{11}~\mathrm{cm}^{-3}\), \(N_{\mathrm{Rb}} = 2 \times 10^{8}\),
and \(n_{\mathrm{Cs}} = 3 \times 10^{11}~\mathrm{cm}^{3}\),
\(N_{\mathrm{Cs}} = 3 \times 10^{8}\).  The temperature of both
species was \(\sim \! 75~\mu\)K.  The atoms were photoassociated
by a Ti:Sapphire laser with intensity of \(\sim \!
3~\mathrm{kW/m}^{2}\). Its frequency was locked to an \(\Omega
=0^{-}\), \(J^P=1^+\) level, located \(38.02~\mathrm{cm}^{-1}\)
below the \(\mathrm{Rb}~ 5S_{1/2} (F=2) + \mathrm{Cs}~ 6P_{1/2}
(F=3)\) atomic asymptote.

All other laser light consisted of pulses with \(\sim \! 5\)~ns
duration and \(\sim \! 3\)~mm diam.  The pump pulse was generated
from a tunable dye laser operating from
18100--18600~\(\mathrm{cm}^{-1}\) at a 10~Hz repetition rate with
a spectral linewidth of \(\sim \! 0.05~\mathrm{cm}^{-1}\).  This
output was sent through a \(\mathrm{H}_{2}\) Raman cell; the
second Stokes order (down-shifted by \(2 \times
4155.25~\mathrm{cm}^{1}\)) was separated to form the pump pulse,
with a typical intensity of \(20~\mathrm{J/m}^{2}\) and frequency
from 9800–-10300~\(\mathrm{cm}^{-1}\).  The dump and re-excitation
pulses were generated using an additional dye laser (the ``red''
laser) operating from 13500-–14000~\(\mathrm{cm}^{-1}\) with a
spectral linewidth of \(\sim \! 0.2~\mathrm{cm}^{-1}\); each had a
typical intensity of \(3~\mathrm{J/m}^{2}\).  The ionizing pulse
(at 532~nm) was derived from the second harmonic of the Nd:YAG
laser used to pump the dye lasers, and had an intensity of
\(100~\mathrm{J/m}^{2}\).

The red and ionizing beams were spatially combined using a
dichroic mirror with the ionizing pulse propagating 4~ns behind
the red pulse.  These combined beams were sent through a beam
splitter.  In one half of the split two-color beam, the optical
path length was adjusted such that the red dump pulse reached the
molecules 7~ns after the pump pulse.  The other half of the
two-color beam was sent through a 22~m long multimode optical
fiber; with this delay, the red re-excitation pulse arrived 110~ns
after the dump pulse.

Laser frequencies were measured using a wavelength meter with
\(0.05~\mathrm{cm}^{-1}\) accuracy.   Ions were detected using a
channeltron located \(\sim \! 3~\mathrm{cm}\) from the atoms.  The
digitized channeltron current yields a time-of-flight mass
spectrum, as shown in the inset of Fig.~\ref{fig01}.  The
temperature of the molecules in the \(a^{3}\Sigma^{+}\) state was
measured as described in Ref.~\cite{Rangwala02}.  The scattering
of two photons necessary to transfer the population to the
\(X(v=0)\) state should not cause significant heating.

\begin{figure}
\centering
\includegraphics[width=3.4in]{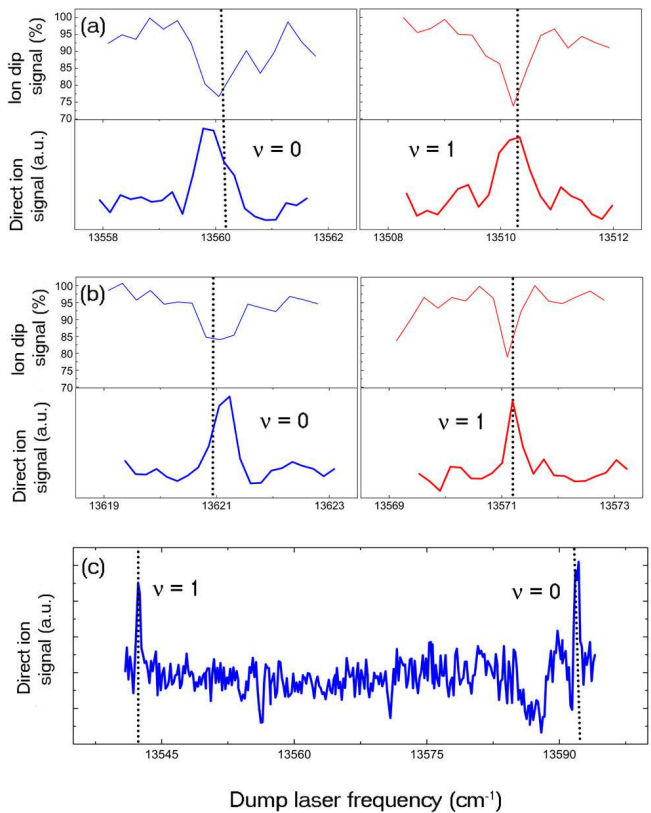}
\caption{(color online) Observation of \(X^{1}\Sigma^{+}(v= 0,1)\)
state molecules. Results are shown for \textit{i} state depletion
(upper) and direct detection (lower) for three consecutive states
\textit{i}, located at energies of (a)
\(9754.26~\mathrm{cm}^{-1}\), (b) \(9814.60~\mathrm{cm}^{-1}\),
and (c) \(9786.10~\mathrm{cm}^{-1}\) above the
\(a^{3}\Sigma^{+}(v=37)\) state.  In (c), the region between the
\(v=0\) and \(v=1\) resonances is shown to have no additional
features for direct detection.  The dotted lines indicate the
predicted dump laser frequency for the desired transition.}
\label{fig02}
\end{figure}

For a particular choice of state \textit{i}, we scanned the
frequency of the red laser through the predicted energy splitting
between the chosen level and the \(X^{1}\Sigma^{+}(v=0,1)\)
states. We observed transfer to the ground states via several of
these levels. Results are shown for three such intermediate states
\textit{i}, consecutive in energy, in Fig.~\ref{fig02}.

The locations of the resonant features are in excellent agreement
with the known \(X(v=0)-(v=1)\) splitting \cite{Fellows99}, and
with our measured splitting of the \textit{i} state levels.  In
Fig.~\ref{fig02}c, a scan over the region between the \(v=0\) and
\(v=1\) resonances shows features only at the predicted locations.
As a cross-check, we also deliberately chose as state \textit{i}
the \(\Omega = 0^-\) component of a mixed
\(c^{3}\Sigma^{+}/b^{3}\Pi\) level.  Here, selection rules
rigorously prevent coupling to the \(X^{1}\Sigma^{+}\) state, and
as expected we observe no evidence for transfer.  Finally, we
determined the strength of the \(i-X\) transitions by increasing
the red laser intensity until the resonant features broadened. The
observed saturation intensities (typically \(\sim \!
4~\mathrm{J/m}^{2}\)) agree qualitatively with those predicted
from calculated FC factors and electronic transition moments
\cite{Bergeman04}.

Presently, we detect \(\sim \! 1\) ion per pulse. This indicates a
production rate of $\sim \! 2 \times 10^2$ UPMs/s, subject to
uncertainties in the channeltron gain and ionization efficiency.
Based on the agreement between observed and predicted rates, this
figure should be correct to within a factor of $\sim 2$. The UPM
production rate is suboptimal for a few reasons. First, MOT
densities were not optimized for this work. Second, the molecules
are untrapped, and most leave the region before our low
repetition-rate lasers can transfer them. Finally, the efficiency
of the SEP is only \(\sim \! 6\)\%. This is determined by the
measured \(\sim \! 25\)\% dump and re-excitation efficiencies and
our inference of a similar pump efficiency (due to the comparable
red and pump laser linewidths). This low value is expected due to
the sparse comb-like spectral structure typical of pulsed lasers
with multiple longitudinal modes \cite{Weber90}, such as that used
here. This structure implies that we would need to broaden the
transitions considerably in order to achieve the theoretical
maximum 25\% transfer efficiency for SEP with our current lasers.

By changing our experimental conditions in straightforward ways,
the method described here can lead to large rates of UPM
production. Thirty-fold higher \(a(v=37)\) production rates could
be achieved by returning to our optimized MOT conditions
\cite{Kerman04}. Using higher repetition rate pulsed lasers and/or
trapping the RbCs molecules could lead to \(\sim \! 100\)\%
addressing of the vibrationally excited molecules.  Finally, using
a stimulated Raman adiabatic passage (STIRAP) technique with
transform-limited laser pulses \cite{He90,Sussman94}, the
\(a(v=37)\) to \(X(v=0)\) transfer efficiency would be \(\sim \!
100\)\%.  With these improvements, we project that the formation
rate of RbCs~\(X(v=0)\) molecules could approach \(3 \times
10^{6}/\mathrm{s}\).

Our $X ^1\Sigma^+$ molecular sample is of high purity with respect
to the vibrational degree of freedom.  State \textit{i} is
dominantly triplet in character and thus the probability of
uncontrolled spontaneous decay to the \(X^{1}\Sigma^{+}\) state
during the SEP step is extremely small. Using our earlier analysis
of the mixed $c/B/b$ state structure \cite{Bergeman04}, we
calculated branching ratios for decay of the \textit{i} states.
While these calculations are only qualitative (due our incomplete
knowledge of the relevant state wavefunctions), they indicate that
the total population in all other vibrational levels of the
\(X^{1}\Sigma^{+}\) state is only \(\sim \! 1\)\% of the
population driven into the \(v=0\) level. (Of course, there
remains a substantial background of \(a^{3}\Sigma^{+}\)
molecules.)

The rotational and hyperfine state distribution of the \(X(v=0)\)
molecules is determined by selection rules and by the spectral
resolution of our lasers.  Hyperfine structure (hfs) is unresolved
in all stages of the process, and hence the nuclear spin degrees
of freedom are completely unconstrained.  However, the initial PA
step selects a level with well-defined rotation/parity quantum
numbers $J^P = 1^+$.  The three subsequent photons involved in the
transfer to the $X$ state (one in spontaneous emission and two in
the SEP process) must leave the molecule in a state of odd parity,
and can add up to 3 units of angular momentum.  Noting that $P =
(-1)^J$ for the $X ^1\Sigma^+$ state, this suggests that only
$J=1,3$ could be populated in $X(v=0)$. However, in principle the
molecule can also acquire rotational angular momentum from the
nuclear spins, if the coupling between hfs and rotation is
sufficiently strong \cite{hfscomment}.  This can add up to $I_{Rb}
+ I_{Cs} = 6$ units of rotation (here $I_x$ is the nuclear spin of
species $x$).  If this coupling is present, the range of
rotational states populated could increase to $J = 1,3,5,7,9$.

We derive an experimental bound on the spread of populated
rotational levels from the absence of large spectral shifts of the
observed resonances from their predicted \(J=0\) locations. Our
observations deviate by no more than \(0.3~\mathrm{cm}^{-1}\) from
these predictions, which have an uncertainty of \(\pm
0.6~\mathrm{cm}^{-1}\) \cite{Bergeman04}. Using the known
rotational constant \(B_e=0.017~\mathrm{cm}^{-1}\)
\cite{Fellows99}, we conclude that at most, \(J=1,3,5,\) and \(7\)
are populated.

The spectral width of the dump and re-excitation resonances
suggests an even narrower rotational distribution.  A typical
linewidth of \(0.18~\mathrm{cm}^{-1}\) (obtained by subtracting,
in quadrature, our laser linewidth from the measured resonance
width) agrees well with the predicted \(0.17~\mathrm{cm}^{-1}\)
splitting between the \(J=1\) and 3 levels.  This suggests that
the hfs/rotation coupling may be absent (or weak), and that indeed
only two rotational levels (\(J=1,3\)) are populated in our
experiments.

In future work, we expect to produce molecules in a single
rotational level by using transform-limited laser pulses capable
of resolving rotational structure.  (Such narrowband lasers are
required in any case for efficient STIRAP transfer to $X(v=0)$).
State selection of hfs may also be possible by the use of
spin-polarized atoms plus judicious choices of laser polarizations
and hfs-resolved transitions for PA and state transfer. This would
also further increase the rate of \(X(v=0)\) molecule production,
since currently the spread of population among hfs sublevels in
the \(a(v=37)\) state limits the fraction available for transfer
to the \textit{i} state.

Finally, we point out that this method for UPM production is quite
general.  Mixing of the \(c^{3}\Sigma^{+}\) and \(B^{1}\Pi\)
levels, as well as the favorable location of the potential curves,
is present in all bi-alkali dimers \cite{Stwalley04}. Although
such spin-orbit mixing increases with the mass of the molecules
\cite{Lefebvre-Brion86}, it may be sufficiently large even for the
lightest heteronuclear bi-alkali (LiNa).  Also, the SEP method is
fairly insensitive to the initial vibrational state and completely
insensitive to molecular temperature.  Thus, higher phase space
densities of UPMs could be produced by photoassociating colder
atoms or by starting with molecules formed by Feshbach resonance
in a single \(a^{3}\Sigma^{+}\) level \cite{Inouye04}.

In summary, we have produced ultracold polar RbCs molecules in
their ground vibronic state.  The optical transfer technique used
here should be applicable to other bi-alkali molecules.
Translational temperatures limited only by atomic cooling methods
should be achievable.  The rotational distributions are limited
only by laser spectral linewidth; with commonly available
technology, population of a single rovibronic state with high
purity should be possible.  This opens a route to the study and
manipulation of polar molecules in the ultracold regime.

We thank A.J. Kerman for crucial contributions to earlier stages
of this work and R.C. Hilborn for the loan of essential equipment.
We acknowledge support at Yale from NSF Grant DMR0325580, the
David and Lucile Packard Foundation, and the W.M. Keck Foundation;
and at Stony Brook from NSF grant PHY0354211 and the U.S. Office
of Naval Research.


\begin{thebibliography}{99}
\bibitem{DeMille02} D. DeMille, Phys. Rev. Lett. \textbf{88}, 067901 (2002).
\bibitem{Baranov02} M. A. Baranov \textit{et al.}, Phys. Rev. A \textbf{66}, 013606 (2002); K. Goral, L. Santos, and M. Lewenstein, Phys. Rev. Lett.
\textbf{88}, 170406 (2002).
\bibitem{Bodo02} E. Bodo, F. A. Gianturco, and A. Dalgarno,  J. Chem. Phys. \textbf{116}, 9222 (2002).
\bibitem{Kozlov95} M. Kozlov and L. Labzowsky, J. Phys. B \textbf{28}, 1933 (1995); J. J. Hudson \textit{et al.}, Phys. Rev. Lett. \textbf{89}, 023003
(2002); D. DeMille, Bull. Am. Phys. Soc. \textbf{49}, 97 (2004).
\bibitem{Weinstein98} J.D. Weinstein et al., Nature (London) \textbf{395}, 148 (1998).
\bibitem{Bethlem00} H. L. Bethlem \textit{et al.}, Nature (London) \textbf{406}, 491 (2000).
\bibitem{Ellioff03} M. S. Ellioff, J. J. Valentini, and D. W. Chandler, Science \textbf{302}, 1940 (2003).
\bibitem{Rangwala02} S. A. Rangwala \textit{et al.}, Phys. Rev. A \textbf{67}, 043406 (2002).
\bibitem{Kerman04} A. J. Kerman \textit{et al.}, Phys. Rev. Lett. \textbf{92}, 153001 (2004).
\bibitem{Mancini04} M. W. Mancini \textit{et al.}, Phys. Rev. Lett. \textbf{92}, 133203 (2004).
\bibitem{Haimberger04} C. Haimberger \textit{et al.}, Phys. Rev. A. \textbf{70}, 021402(R) (2004).
\bibitem{Wang04} D. Wang \textit{et al.}, Phys. Rev. Lett. \textbf{93}, 243005 (2004).
\bibitem{Inouye04} S. Inouye \textit{et al.}, Phys. Rev. Lett. \textbf{93}, 183201 (2004); C. A. Stan \textit{et al.}, Phys. Rev. Lett. \textbf{93}, 143001
(2004).
\bibitem{Kotochigova03} S. Kotochigova, P. S. Julienne, and E. Tiesinga, Phys. Rev. A \textbf{68}, 022501 (2003).
\bibitem{Yurovsky00} V. A. Yurovsky \textit{et al.}, Phys. Rev. A \textbf{62}, 043605 (2000).
\bibitem{Mukaiyama04} T. Mukaiyama \textit{et al.}, Phys. Rev. Lett. \textbf{92}, 180402 (2004).
\bibitem{Damski03}  B. Damski \textit{et al.}, Phys. Rev. Lett. \textbf{90}, 110401 (2003).
\bibitem{Bergeman04} T. Bergeman \textit{et al.}, Eur. Phys. J. D \textbf{31}, 179 (2004).
\bibitem{Stwalley04} W. C. Stwalley, Eur. Phys. J. D \textbf{31}, 221 (2004).
\bibitem{Kotochigova04} S. Kotochigova, E. Tiesinga, and P. S. Julienne, Eur. Phys. J. D \textbf{31}, 189 (2004).
\bibitem{IgelMann86} G. Igel-Mann \textit{et al.},  J. Chem. Phys. \textbf{84}, 5007 (1986); S. Kotochigova, private communication.
\bibitem{Thorsheim87} H. R. Thorsheim, J. Weiner, and P. S. Julienne, Phys. Rev. Lett. \textbf{58}, 2420 (1987); for a review, see: F. Masnou-Seeuws and P. Pillet, Adv. At. Mol. Opt. Phys. \textbf{47}, 53
(2001).
\bibitem{Cooper81} D. E. Cooper, C. M. Klimcak, and J. E. Wessel, Phys. Rev. Lett. \textbf{46}, 324 (1981).
\bibitem{Fellows99} C. E. Fellows \textit{et al.}, J. Mol. Spectrosc. \textbf{197}, 19 (1999).
\bibitem{Kerman04b} A. J. Kerman \textit{et al.}, Phys. Rev. Lett. \textbf{92}, 033004 (2004).
\bibitem{Ketterle93} W. Ketterle \textit{et al.}, Phys. Rev. Lett. \textbf{70}, 2253 (1993).
\bibitem{Anderson94} M. H. Anderson \textit{et al.}, Phys. Rev. A \textbf{50}, R3597 (1994).
\bibitem{Weber90} Th. Weber, E. Riedle, and H. J. Neusser, J. Opt. Soc. Am. B \textbf{7}, 1875 (1990).
\bibitem{He90} G. He \textit{et al.}, J. Opt. Soc. Am. B \textbf{7}, 1960 (1990).
\bibitem{Sussman94} R. Sussman, R. Neuhauser, and H. J. Neusser, J. Chem. Phys. \textbf{100}, 4784 (1994).
\bibitem{hfscomment} Strong hfs/rotation coupling requires that hfs be large
compared to rotational splittings.  This cannot occur in the $X
^1\Sigma^+$ or $\Omega = 0^-$ PA states, where the small hfs
\cite{Fellows99, Kerman04b} arises only from electric quadrupole
effects.  However, strong hfs/rotation coupling could occur in the
$a ^3\Sigma^+$ and $c ^3\Sigma^+$ states, where magnetic dipole
hfs is present \cite{Townes55}. Crude estimates indicate that hfs
might exceed rotational splittings for low-$J$ levels in these
states.
\bibitem{Townes55} C.H. Townes and A.L. Schawlow, \textit{Microwave Spectroscopy} (McGraw-Hill, New York, 1955).
\bibitem{Lefebvre-Brion86} H. Lefebvre-Brion and R. W. Field, \textit{The Spectra and Dynamics of Diatomic Molecules} (Elsevier, San Diego, 2004).
\end{thebibliography}
\end{document}